\documentclass[twocolumn,showpacs,aps,epsfig]{revtex4}

\usepackage{graphicx}
\usepackage{epstopdf}
\usepackage{latexsym}
\usepackage{amssymb}

\usepackage[center]{subfigure}

\begin{document}

%%%%%%%%%%%%%%%%%%%%%%%%%%%%%%%%%%%%%%%%%%%%%%%%%%%%%%%%%%%%%%%
 \newcommand{\bq}{\begin{equation}}
 \newcommand{\eq}{\end{equation}}
 \newcommand{\bqn}{\begin{eqnarray}}
 \newcommand{\eqn}{\end{eqnarray}}
 \newcommand{\nb}{\nonumber}
 \newcommand{\lb}{\label}
\newcommand{\PRL}{Phys. Rev. Lett.}
\newcommand{\PL}{Phys. Lett.}
\newcommand{\PR}{Phys. Rev.}
\newcommand{\CQG}{Class. Quantum Grav.}
 %%%%%%%%%%%%%%%%%%%%%%%%%%%%%%%%%%%%%%%%%%%%%%%%%%%%%%%%%%%%%%%

\title{An Alternative Approach for General Covariant Ho\v rava-Lifshitz Gravity and Matter Coupling} 

\author{Alan M. ds Silva}
\email{amsilva@fma.if.usp.br}

%\author{}
%\email{}

\affiliation{Instituto de F\'isica, Universidade de S\~ao Paulo\\
C.P.66.318, CEP 05315-970, S\~ao Paulo, Brazil}

\date{\today}

\begin{abstract}

Recently, Ho\v rava and Melby-Thompson proposed in \cite{Horava 2.0} a nonrelativistic gravity theory with extended gauge symmetry that is free of the spin-0 graviton. We propose a minimal substitution recipe to implement this extended gauge symmetry which reproduce the results obtained by them. Our prescription has the advantage of being manifestly gauge invariant and immediately generalizable to other fields, like matter. We briefly discuss the coupling of gravity with scalar and vector fields found by our method. We show also that the extended gauge invariance in gravity does not force the value of $\lambda$ to be $\lambda=1$ as claimed in \cite{Horava 2.0}. However, the spin-0 graviton is eliminated even for general $\lambda$.

\end{abstract}
\pacs{04.60.-m; 98.80.Qc;98.80.-k}
%\pacs{04.60.-m; 98.80.Cq; 98.80.-k; 98.80.Bp}

\maketitle

\section{Introduction}

In  \cite{Membranes, Horava 1.0}, Ho\v rava proposed a power-counting ultraviolet (UV) renomalizable quantum gravity theory which has attracted a great interest. The theory was inspired by the Lifshitz scalar \cite{lipstick} and has often been called Ho\v rava-Lifshitz (HL) gravity. This theory exhibits an improved UV behavior and it is expected that the theory may be compatible with available experimental data at its infrared (IR) limit. However, deviations from standard general relativity (GR) are possible.

Ho\v rava theory is constructed on the basic assumption of anisotropic scaling between space and time, \emph{i.e.},
\begin{equation}
\textbf{x} \rightarrow b \textbf{x}, \qquad t \rightarrow b^{z} t
\label{scaling}
\end{equation}
\noindent
where $z$ is the dynamical critical exponent. The scaling dimension of an operator $\phi$ is defined by its transformation under (\ref{scaling}). If $\phi \rightarrow b^{-s}\phi$, then $[\phi]=s$ is the scaling dimension of $\phi$. A power-counting renormalizable $(D+1)$-dimensional gravity theory requires $z=D$ (see \cite{quantum regulator}), what will be always assumed in this paper.

In a theory with anisotropic scaling, time and space are fundamentally distinct, then it is natural to use the ADM formalism which splits spacetime into space slices and time. In the ADM formalism the spacetime metric is decomposed as:
\begin{equation}
ds^2 = -N^2 dt^2 + g_{ij}(dx^i +N^i dt)(dx^j + N^j dt).\label{metrica ADM}
\end{equation}
\noindent

The anisotropy between space and time also implies that HL gravity does not share GR's general diffeormorphism invariance, the local symmetries being restricted to
\begin{equation}
\delta t = f(t), \qquad \delta x^i = \xi^i(t, x^j), \label{DiffMF}
\end{equation}
\noindent
which are the foliation preserving diffeomorphisms, Diff$(\mathcal{M}, \mathcal{F})$, where $\mathcal{M}$ is the spacetime manifold, provided with a preferred foliation structure $\mathcal{F}$. In a simpler way, we will just consider $\mathcal{M}= \textbf{R} \times \Sigma$, where $\Sigma$ is a $D$-dimensional manifold corresponding to spacial slices of constant $t$. Under (\ref{DiffMF}), the ADM fields $N$, $N^i$ and $g_{ij}$ transforms as:
\begin{eqnarray}
\delta g_{ij}&=&\partial _i \xi ^k g_{jk}+\partial _j \xi ^k g_{ik}+\xi ^k \partial _k g_{ij}+f\dot{g}_{ij}, \nonumber\\
\delta N_i&=&\partial _i \xi ^j N_j+\xi ^j \partial _j Ni+ \xi ^j g_{ij}+\dot{f}N_i+f\dot{N}_i\nonumber\\
\delta N&=&\xi ^j \partial _j N+\dot{f}N+f\dot{N}. \label{transf. GL}
\end{eqnarray}

If we restrict $N$ to be $N=N(t)$ the theory is called projectable. Otherwise, for $N=N(t, x^i)$, we will have the nonprojectable HL theory. In the rest of this paper we will always assume projectability in HL theory.

The action of a $D+1$-dimensional HL theory has the form:
\begin{equation}
S=S_K - S_V,
\end{equation}    
\noindent
where  
\begin{eqnarray}
S_K &=& \int{dtd^D x\sqrt{g} N \left[K_{ij}G^{ijkl}K^{kl}\right]}=\nonumber\\
&=&\int{dtd^D x\sqrt{g} N \left[K_{ij} K^{ij} - \lambda K^2\right]}\label{cinetico}
\end{eqnarray} 
\noindent
is the kinetic term, which contains the time derivatives, with
\begin{equation}
G^{ijkl}=\frac{1}{2}\left( g^{ik}g^{jl}+g^{il}g^{jk}\right)-\lambda g^{ij}g^{kl}
\end{equation}
\noindent
being a generalized DeWitt metric. The $\lambda$ parameter comes from the absence of spacetime diffeomorphism symmetry due to anisotropic scaling. The kinetic term in GR, written in the ADM formalism, has the same form of (\ref{cinetico}), but with $\lambda=1$. A mechanism to make $\lambda$ be close to 1 in HL theory, at least at the IR limit, is necessary to match observational constraints\cite{dutta}. Such a mechanism is still unknown, but a possible solution is that $\lambda=1$ may be a IR fixed point of the RG flow of the theory. Another plausible way to fix $\lambda =1$ could be the assumption of extra local symmetries which impose this value to $\lambda$. 

The tensor $K_{ij}$ is the extrinsic curvature of spatial slices defined by:
\begin{equation}
K_{ij}=\frac{1}{2N}\left(\dot{g}_{ij}-\nabla _i N_j - \nabla _j N_i \right)
\end{equation}
\noindent
where a dot indicates the time derivative, $\nabla$ is the covariant derivative on the spatial slice $\Sigma$, whose metric $g_{ij}$ is used to raise and lower indices. 

The potential term $S_V$ is defined by:
\begin{equation}
S_V = \int{dtd^D x \sqrt{g} N \mathcal{V}(g_{ij})}
\end{equation}
\noindent
where $\mathcal{V}$ is a scalar operator built with the spacial metric and its spacial derivatives, with $[\mathcal{V}]=2z$. In order to recover the relativistic $z=1$ at long distances, $\mathcal{V}$ should contain the relevant operators \cite{Horava 1.0, Sotiriou}:
\begin{equation}
\mathcal{V}_{IR}=-\chi^4 (R-2\Lambda)
\end{equation}
\noindent
where $\chi$ is a fundamental momentum scale of the theory. The most general $\mathcal{V}$ in projectable HL theory contains all operators with six spatial derivatives of the metric or less, and is given in \cite{Sotiriou}.

Projectable HL gravity has been intensely explored since the original proposal of \cite{Membranes,Horava 1.0} and some issues presented there remain unresolved, as the lack of a known way to make $\lambda=1$ at IR limit and the presence of an extra degree of freedom in gravity, related with the reduced local symmetries of the theory, which has been called spin-0 graviton or scalar graviton, whose behavior still has not been completely understood, but may ruin the obtention of a phenomenologically viable theory of gravity at IR limit (see a recent discussion of the scalar graviton issue in \cite{Shinji}).

\section{General Covariant HL Gravity}

In order to solve such issues, in \cite{Horava 2.0}, the gauge symmetry of HL gravity was extended by a local $U(1)$ symmetry. The fields transform under this $U(1)$ symmetry as
\begin{eqnarray}
\delta _{\alpha} N &=& \delta _{\alpha} g_{ij} = 0,\nonumber\\ 
\delta _{\alpha} N_i &=& N \nabla _ i \alpha.
\label{transf. U1}
\end{eqnarray}

This extension was motivated by the fact that linearized HL gravity around flat spacetime has a global $U(1)_{\Sigma}$ symmetry for $\lambda=1$,\emph{ i.e.}, it is invariant under the flat spacetime version of (\ref{transf. U1}) if $\dot{\alpha}=0$. This global symmetry can be interpreted as residual gauge symmetry after a gauge fixing. By including a coupling with a gauge field $A$, which transforms as $\delta _{\alpha}A=\dot{\alpha}$ we can make the theory invariant under (\ref{transf. U1}) for time-dependent $\alpha$, thus, extending the gauge group to $U(1) \times$ Diff$(\mathcal{M}, \mathcal{F})$ (details in  \cite{Membranes, Horava 2.0}).

The generalization of this result to non-linear theory is not straighfoward for $D\geq 3$, because even at $\lambda=1$, the theory does not present any extra symmetry. The HL action transforms, for $\lambda =1$, as
\begin{eqnarray}
\delta _{\alpha} S&=&-\int{dt d^D x \sqrt{g} \alpha \left(R^{ij} - \frac{1}{2} R g^{ij} \right)}(\dot{g}_{ij}-\nabla _{(i} N_{j)})\nonumber\\
 &+&\int{dt d^D x \sqrt{g} (\dot{\alpha}- N^i \nabla _i \alpha) R}.
\label{delta S}
\end{eqnarray}
This expression does not vanish, even under the condition of vanishing time derivative, which is $\dot{\alpha} - N^i \nabla _i \alpha =0 $ in curved spacetime. Thus, there is no extra invariance at $\lambda=1$.
The exception is the $D=2$ case, because the first term of (\ref{delta S}) vanishes identically in two dimensions. Then, one introduces the gauge field $A$, which transforms as
\begin{equation}
\delta _{\alpha} A = \dot{\alpha}- N^j \nabla _j \alpha
\label{delta A}
\end{equation}
\noindent
and has scaling dimension $[A]=2z-2$. The addition of the coupling term
\begin{equation}
S_A= - \int{dtd^D x \sqrt{g} A R}
\end{equation}
makes the theory invariant under ($\ref{transf. U1}$) for general $\alpha$.

The solution given to acomplish the symmetry extension for $D \geq 3$ was the coupling of an auxiliary scalar field $\nu$, called the newtonian prepotential, which transforms as $\delta _{\alpha} \nu = \alpha$ and has scaling dimension $z-2$.
An action invariant under global $U(1)_{\Sigma}$ was constructed by coupling the $\nu$-field in a convenient way
\begin{eqnarray}
S&=& \int{dtd^D x \sqrt{g} N\left[K_{ij}K^{ij} - K^2 - \mathcal{V}(g_{ij})\right]}\nonumber\\
 &+&\int{dtd^D x \sqrt{g} N \nu \Theta ^{ij}\left(2 K_{ij} + \nabla _i \nabla _j \nu\right)}
\label{acao do horava}
\end{eqnarray}
\noindent
with
\begin {equation}
\Theta ^{ij}=R^{ij}-\frac{1}{2}g^{ij}R+ \Omega g^{ij}.
\end{equation}
Action (\ref{acao do horava}) transforms under (\ref{transf. U1}) as
\begin{equation}
\Delta _{\alpha} S= \int{dt d^D x \sqrt{g} (\dot{\alpha} - N^j \nabla _j \alpha)(R-2\Omega)}.
\end{equation}
Finally, the $A$-term can be added to the action (\ref{acao do horava}):
\begin{equation}
S_A = - \int{dt d^D x \sqrt{g} A (R-2\Omega)}
\end{equation}
and we obtain a $(D+1)$-dimensional nonrelativistic gravity theory whose gauge group of symmetry is $U(1) \times$ Diff$(\mathcal{M}, \mathcal{F})$. Such a theory is called general covariant because the number of local symmetries is $D+1$ per spacetime point, as in general relativity. This extra symmetry has a simple geometrical interpretation as the subleading order of the expansion of spacetime diffeomorphisms in powers of $1/c$. In the Newtonian aproximation, the gauge field $A$ plays the role of the newtonian potential (details in \cite{Horava 2.0}).

The variation of $S$ with respect to $A$ gives 
\begin{equation}
R-2\Omega=0, \label{A equation}
\end{equation}
\noindent
which is a constraint equation. This extra constraint provides that the theory is free of spin-0 graviton and has only the two usual tensor degree of freedom propagating.

\section{U(1) symmetry by a minimal substitution}

One of the issues of general covariant HL gravity is that it is not evident how to couple it with matter. Our goal in this section is to derive the $(D+1)$-dimensional general covariant HL theory described above in a manifestly gauge invariant way, by using a "`minimal substitution" recipe. The advantage of having a straightfoward method like this one is to be able to apply it to extend the symmetry of any Diff$(\mathcal{M}, \mathcal{F})$ invariant action, like those proposed in \cite{Kiritsis e Kofinas} for nonrelativistic scalar and vector fields. Extending the gauge symmetry of matter fields to construct $U(1) \times$ Diff$(\mathcal{M}, \mathcal{F})$ invariant actions will give us naturally the answer of how can one couple matter with the new gravitational fields, $A$ and $\nu$.

Given an action $S[N, N_i, g_{ij}, \psi_n]$, Diff$(\mathcal{M}, \mathcal{F})$ invariant, where $\psi_n$ are arbitrary fields, we can introduce the coupling with $A$ and $\nu$ to obtain the $U(1) \times$ Diff$(\mathcal{M}, \mathcal{F})$ invariant action $\hat{S}[N, N_i, g_{ij}, \psi_n]$, by adopting the following prescription:
\begin{eqnarray}
\hat{S} &=& S[N, N_i - N\nabla _i \nu, g_{ij}, \psi_n]+\nonumber\\
&+&\int{dtd^{D}x\sqrt{g}Z(\psi_n, g_{ij})(A-a)},
\end{eqnarray}
\noindent
where $Z$ is the most general scalar operator under the full $U(1) \times$ Diff$(\mathcal{M}, \mathcal{F})$ , with $[Z]=2$, built with fields and metric. This is necessary because the lagrangian density must have scaling dimension $2z$ in HL theory. The field $a= \dot{\nu}-N^j \nabla _j \nu + \frac{N}{2} \nabla ^j \nu \nabla _j \nu$ transforms under $U(1)$ as $\delta _{\alpha} a = \dot{\alpha} - N^i \nabla _i \alpha $. Thus, both pieces of the new action are manifestly invariant under (\ref{transf. U1}).

We will use this prescription to reproduce the action (\ref{acao do horava}), showing the equivalence of our approach and the procedure used in \cite{Horava 2.0}. Under the transformation $N_j \rightarrow N_j - N \nabla _j \nu $, we have $K_{ij} \rightarrow K_{ij} + \nabla _i \nabla _j \nu $. We obtain, for $\lambda=1$:
\begin{eqnarray}
S&=& \int{dtd^D x \sqrt{g} N(K_{ij} + \nabla _i \nabla _j \nu) G^{ijkl}(K_{kl} + \nabla _k \nabla _l \nu)} \nonumber\\
&-& V(g_{ij}) + c(R-2 \Omega )(A-a).
\label{minha acao}
\end{eqnarray}
\noindent
where $G^{ijkl}=\frac{1}{2}\left( g^{ik}g^{jl}+g^{il}g^{jk}\right)- g^{ij}g^{kl}$ is the DeWitt metric.

Note that we can choose the constant $c$ in the $A-a$ term by absorbing a constant in $A$ and $a$ with no loss of generality. We choose $c=-1$ to match (\ref{acao do horava}) and keep the geometrical interpretation of the $A$-field. The $\nu$-dependent term in the action is
\begin{eqnarray}
S_{\nu}&=&\int{dtd^D x \sqrt{g}N [2K_{ij}G^{ijkl}\nabla _k \nabla _l} \nu\nonumber\\
 &+& (\nabla _i \nabla _j \nu)G^{ijkl}(\nabla _k \nabla _l \nu) +(R-2\Omega)a] ,
\end{eqnarray}
\noindent
which can be proven to be identical to
\begin{equation}
\int{dtd^D x \sqrt{g}N \left[\nu \Theta ^{ij}(2 K_{ij} + \nabla _i \nabla _j \nu)\right]}
\end{equation}
\noindent
found in \cite{Horava 2.0}.

Note that the fixing $\lambda=1$ is clearly not necessary in our derivation and we did it just to show that our method can reproduce (\ref{acao do horava}). Hence, we can restablish the free $\lambda$ in nonrelativistic gravity theory by just adding in  the $\lambda$-term below:
\begin{equation}
S_{\lambda}=\int{dtd^D x \sqrt{g} N (1- \lambda)(K + \Delta \nu)^2}.\label{lambda-term}
\end{equation}
\noindent
or, equivalently, by using the generalized DeWitt metric:
\begin{equation}
G^{ijkl}=\frac{1}{2}\left( g^{ik}g^{jl}+g^{il}g^{jk}\right)-\lambda g^{ij}g^{kl}
\label{de witt geral}
\end{equation}
\noindent
as done in previous HL type theories. This is in disagrement with the claim made in \cite{Horava 2.0} that the extended gauge symmetry forces $\lambda$ to be one. 

\section{Linear Theory with Free $\lambda$}

One of the major features of general covariant HL theory is being free of the extra degree of freedom that used to appear in the scalar spectrum of nonrelativistic gravity\footnote{The absence of the scalar graviton in Minkowski background using the most general potential term and $\lambda=1$ was recently shown in \cite{chineses}.}. Now that we have reintroduced the $\lambda$-dependence in general covariant HL theory, we should verify if for $\lambda \neq 1$ the theory is still scalar graviton free.
Following the steps of \cite{Horava 2.0} we will linearize the theory around a ground state solution in the particular case when the detailed balance condition is satisfied, for simplicity. The potential term takes the form
\begin{equation}
V=\frac{1}{4}\mathcal{G}_{ijkl} \frac{\delta W}{\delta g_{ij}}\frac{\delta W}{\delta g_{kl}},
\end{equation}
\noindent
where $\mathcal{G}_{ijkl}$ is the inverse of DeWitt metric (\ref{de witt geral}) and
\begin{equation}
W = \int{d^D x \sqrt{g}(R-2\Lambda _W)}.
\end{equation}
\noindent
%The potencial obtained is:
%\begin{equation}
%\mathcal{V}(g_{ij})=\frac{1}{4}(R^{ij}-\frac{1}{2}R g^{ij}+ \Lambda_W g^{ij}) \mathcal{G}_{ijkl}(R^{kl}-\frac{1}{2}R g^{kl}+ \Lambda_W g^{kl}).
%\end{equation}

Our ground state solution is given by
\begin{equation}
g_ij= \hat{g}_{ij}, \qquad N=1, \qquad N_i=0, \qquad, \nu =0, \qquad A=0,
\end{equation}
\noindent
where $\hat{g}_{ij}$ is a time independent and maximally symmetric spatial metric which solves the equation of motion of $W$:
\begin{equation}
R_{ij}-\frac{1}{2}R g_{ij} + \Lambda _W g_{ij}=0,
\end{equation}
\noindent
with
\begin{equation}
\Omega=\frac{D}{D-2} \Lambda _W,
\end{equation}
\noindent
such that $\hat{R}_{ij} = \frac{2 \Omega}{D}$ and $\hat{R}=2\Omega$. 

We need only to search the $\lambda$ effects in the scalar sector of perturbations, because the $\lambda$-term (\ref{lambda-term}) is constructed with the trace of the extrinsic curvature tensor, which receives contributions of scalar perturbations only. The vector and tensor perturbation will behave as the result obtained in \cite{Horava 2.0}.

The most general scalar perturbations are
\begin{equation}
h_{ij} = -2 \psi \hat{g}_{ij} + \hat{\nabla}_i \hat{\nabla} _j E, \qquad n_i=\hat{\nabla} _i B,
\end{equation}
\noindent
where $E$ satisfies $\hat{\Delta} E=0$. We can make $E=0$ by a gauge choice. In this gauge, the $\nu$ equation of motion is:
\begin{equation}
\left(\frac{2 \Omega}{D}\hat{\Delta} +(1-\lambda) \hat{\Delta} ^2\right) (\nu-B) +D(\lambda -1) \hat{\Delta}\dot{\psi} - 2\Omega \dot{\psi}=0, \label{1}
\end{equation}
\noindent
and the momentum constraint:
\begin{equation}
\left(\frac{2 \Omega}{D}\hat{\Delta} +(1-\lambda) \hat{\Delta} ^2\right) (\nu-B)+(\lambda D-1)\hat{\Delta}\dot{\psi} + 2 \Omega \hat{\Delta} \nu =0. \label{2}
\end{equation}

Substracting (\ref{2}) from (\ref{1}), we obtain
\begin{equation}
- (D-1) \hat{\Delta} \dot{\psi} - 2 \Omega \dot{\psi} - 2 \Omega \hat{\Delta} \nu=0.
\end{equation}

Aplying the constraint (\ref{A equation}), which at first order in perturbations is
\begin{equation}
(D-1)\hat{\Delta}\psi + 2\Omega \psi =0
\label{vinculo de gauss}
\end{equation}
\noindent
we obtain
\begin{equation}
\hat{\Delta}\nu =0.
\label{delta nu}
\end{equation}

Equation (\ref{vinculo de gauss}) shows that $\psi$ is not propagating and implies, with (\ref{delta nu}) and (\ref{1}),
\begin{equation}
\hat{\Delta}\left[\frac{2 \Omega}{D}B + (1-\lambda) \hat{\Delta}B\right]=0
\end{equation}
\noindent
which shows that $B$ is not propagating as well.

Hence, as in the case $\lambda =1$, the theory is free of scalar graviton even for general $\lambda$. This result supports the claim, also made in \cite{Horava 2.0}, that the extended $U(1) \times$ Diff$(\mathcal{M}, \mathcal{F})$ symmetry eliminates the extra physical degree of freedom in HL theory.

\section{Matter Coupling}

Since we can reproduce the general covariant HL gravity theory by just including the $U(1)$ gauge symmetry in its action by our minimal substitution recipe, we are able to couple matter fields with gravity by the same method.

We will specialize to the nonrelativistic scalar field in three dimensions, with $z=3$, proposed in \cite{Kiritsis e Kofinas}, to show the application of the prescription we proposed. We find:
\begin{eqnarray}
S&=& \int{dtd^3x \sqrt{g}N \left[\frac{1}{2N^2}\left(\dot{\varphi}-(N^i-N \nabla ^i \nu) \partial _i \varphi \right)^2 - F\right]}\nonumber\\
 &+&\int{dtd^3x \sqrt{g}(c_1(\varphi)\Delta \varphi +c_2(\varphi)\nabla ^i \varphi \nabla _i \varphi)(A-a)}
\label{Sphi}
\end{eqnarray}

We can define a new covariant time derivative operator by
\begin{equation}
D_t = \partial _t - \hat{N}^i \nabla _i ,\qquad \hat{N}^i = N^i - N \nabla ^i \nu,
\end{equation} 
\noindent
and the equation of motion of a scalar field takes the form:
\begin{eqnarray}
&-&\frac{1}{N}D_t \left(\frac{1}{N} D_t \varphi \right)-(K + \Delta \nu)\frac{1}{N}D_t \varphi \nonumber\\
&+& c_{\varphi}^2 \Delta \varphi + I(\varphi, A-a) + ... =0
\end{eqnarray}
\noindent
where we have ommited terms which contain more than two spacial derivatives and the potential $V(\varphi)$. The coupling term $I(\varphi, A-a)$ is
\begin{eqnarray}
I&=& \left[\Delta \varphi \left(2 c^{\prime}_1 -2c_2 \right)+\nabla _i \varphi \nabla ^i \varphi \left(c^{\prime \prime}_1 - c^{\prime}_2 \right) \right] (A-a) \nonumber\\
&+& \nabla _i \varphi \left(2 c^{\prime}_1 - 2c_2 \right)\nabla ^i (A-a) +c_1 \Delta (A-a)
\end{eqnarray}

In the special case $c_2=c_1^{\prime}$ it reduces to $I=c_1 \Delta (A-a)$. It is worth noting also that in the absence of gravity, i.e, $g_{ij} = \delta _{ij}$, $N=1$, $N^i = A = \nu =0$, the equation of motion reduces to a wave equation in flat spacetime as needed to recover the emergent Lorentz symmetry.

By the same procedure, we can couple a vector field of mass $M$ with gravity. Using the nonrelativistic action proposed in \cite{Kiritsis e Kofinas} we obtain
\begin{eqnarray}
S&=& \frac{1}{4g^2}\int{dtd^3x \sqrt{g}\left[\frac{2}{N}g^{ij}(F_{oi} -\hat{N}^k F_{ki})(F_{0j}-\hat{N}^l F_{il})\right]}\nonumber \\
&+& \frac{1}{4g^2}\int{dtd^3x \sqrt{g}N\left[\frac{M^2}{N^2}(A_0 - \hat{N}^i A_i)^2 - G[A_i]\right]} \nonumber\\
&+&\int{dtd^3 x \sqrt{g}\kappa B_i B^i (A-a)} 
\label{vetor}
\end{eqnarray}
\noindent
where $\kappa$ can be any function of $A_i A^i $.

We will not go further on the analysis of those fields in the present paper. We just point out that an immediate consequence of those matter couplings is that we can expect differences between general covariant HL theory and GR, even in the IR limit, when they are coupled with matter. This happens because the coupling of matter with the $A$-field is through scaling dimension two operators which affects the IR behavior. The evaluation of such differences will be treated in future work.

\section{Conclusions}

In this paper we have proposed a minimal substitution recipe that reproduces the general covariant HL gravity action proposed in \cite{Horava 2.0} and we have pointed out that the nonrelativistic general covariance $U(1) \times$Diff$(\mathcal{M}, \mathcal{F})$ does not fix $\lambda =1$ in HL theory as was originally stated. Nonrelativistic general covariance indeed forces the $\lambda=1$ in the cases when the $\nu$-field is not necessary to achieve extended gauge invariance, as HL gravity in $2+1$ dimensions and linearized HL gravity in any dimension, because those theories present a global $U(1)_{\Sigma}$ invariance only for $\lambda=1$. The field $\nu$ is introduced to fix the problem of lack of global $U(1)_{\Sigma}$ invariance in HL theory for $D \geq 3$, but it is a "too powerful" tool for the task, because it can also give global $U(1)_{\Sigma}$ invariance to HL theory for any value of $\lambda$, as our approach makes clear. Hence, $\lambda$ still must be seen as a running coupling constant, and HL theory still lacks a mechanism to make it takes its general relativistic value at long distances.

As a consequence of our approach, we have a way to find the coupling of matter with the newly introduced gravitational fields, by extending the gauge invariance of matter action using the same prescription used in HL gravity. We have applied our method to nonrelativistic scalar and vector fields as an example. It is worth noting that an issue pointed out in \cite{Horava 2.0} is still present with the matter couplings we have found. The $A$ equation of motion in the gravity sector gives the constraint $R=2\Omega$. This constraint does not allow a FRW solution with non zero spatial curvature. It was expected that the matter coupling could solve this problem, but the couplings we found are always built with spatial derivatives, which vanishes if we impose the cosmological principle, and the constraint remains the same. Indeed, this should not be surprising, because the $[A]=2z-2$ implies the coupling with scaling dimension two operators. The case of interest in cosmology is $D=3$, which requires $z \geq 3$ by power-counting. Hence, the scaling dimension of any time derivative is at least three.

Finally, we have shown that even if we let $\lambda \neq 1$, the linear theory around a ground state solution indicates that there is no extra degree of freedom propagating, besides  a traceless and transverse tensor. This is a good sign that the IR limit of general covariant HL theory may reproduce some features of linearized general relativity. However, the raised issues of the $\lambda$-dependence and the different interaction between matter and gravity even in the IR physics need further investigation.

\noindent
\textbf{Acknowlegments:} The author is grateful to E. G. M. Ferreira for the valuable discussions and to FAPESP for the financial support.

%%%%%%%%%%%%%%%%%%%%%%%%%%%%%%%%%%%%%%%%%%%%%%%%%%%%%%

\end{document}